\documentclass[twocolumn,usenatbib,useAMS]{mnras}

\PassOptionsToPackage{dvipsnames,svgnames,x11names}{xcolor}
\usepackage{xparse,soul}

\makeatletter
\NewDocumentCommand{\sotwo}{O{red}O{black}+m}
    {%
        \begingroup
        \color{#1}%
        \setul{-.5ex}{.4pt}%
        \def\SOUL@uleverysyllable{%
            \rlap{%
                \color{#2}\the\SOUL@syllable
                \SOUL@setkern\SOUL@charkern}%
            \SOUL@ulunderline{%
                \phantom{\the\SOUL@syllable}}%
        }%
        \ul{#3}%
        \endgroup
    }
\makeatother

\usepackage{graphicx}	
\usepackage{amsmath}	
\usepackage{amssymb}	
\usepackage{multicol} 
\usepackage{multirow}
\usepackage{bm}	
\usepackage{pdflscape}
\usepackage{caption}
\usepackage{subfigure}
\usepackage{siunitx}
\usepackage{color}
\usepackage{lastpage}
\usepackage{xspace}
\usepackage{color}

\newcommand{\kms}{\,{\rm km\,s}^{-1}}

\usepackage[T1]{fontenc}
\usepackage{ae,aecompl}

\usepackage{newtxtext,newtxmath}

\def\msunoh{\,h^{-1}{\rm M}_\odot}
\def\mpcoh{\,h^{-1}{\rm Mpc}}

\title[Rotating kinetic Sunyaev-Zeldovich effect]{A possible detection of gas rotation in galaxy groups via the kinetic Sunyaev-Zeldovich effect }%\JAP{\dots of gas rotation in galaxy groups \dots ?}}
\author[Yang et al.]{Tianyi Yang$^{1,2}$, Yan-Chuan Cai$^{2}$, Weiguang Cui$^{3,4, 2}$, John A. Peacock$^{2}$,  Romeel Dav\'e$^{2,8}$, 
\newauthor Houjun Mo$^{5}$, Huiyuan Wang$^{6}$, and Xiaohu Yang$^{7}$ \\
$^1$Astrophysics Research Institute, Liverpool John Moores University, Liverpool, L3 5RF, UK \\
$^{2}$Institute for Astronomy, University of Edinburgh, Royal Observatory, Blackford Hill, Edinburgh, EH9 3HJ, UK \\
$^3$Departamento de Física Teórica, Módulo 15, Facultad de Ciencias, Universidad Autónoma de Madrid, 28049 Madrid, Spain \\
$^4$Centro de Investigación Avanzada en Física Fundamental (CIAFF), Facultad de Ciencias, Universidad Autónoma de Madrid, 28049 Madrid, Spain\\
$^5$ Department of Astronomy, University of Massachusetts, Amherst MA 01003-9305, USA \\
$^6$Department of Astronomy, University of Science and Technology of China, Hefei, Anhui 230026, China\\
$^{7}$State Key Laboratory of Dark Matter Physics, Tsung-Dao Lee Institute \& School of Physics and Astronomy, Shanghai Jiao Tong University, Shanghai 201210, China\\
$^{8}$Department of Physics and Astronomy, University of the Western Cape, Bellville, Cape Town 7535, South Africa\\
} 

\date{Accepted XXX. Received YYY; in original form ZZZ}

\pubyear{2020}

\begin{document}
\label{firstpage}
\pagerange{\pageref{firstpage}--\pageref{lastpage}}
\maketitle

\begin{abstract} 
We utilise the kinetic Sunyaev-Zeldovich effect (kSZ) to measure the rotation of ionised gas within galaxy groups defined in the SDSS-DR7 galaxy sample, via their dipolar imprint on the cosmic microwave background (CMB). We estimate the direction of the projected angular momentum for each of our selected fast-rotating groups by measuring the redshift dipole of satellite galaxies around their group centres. We find a clear redshift dipole in the stacked data for the selected SDSS groups. We then perform oriented stacking of the Planck CMB temperature map using the group centres and directions of angular momenta. We report a $2.3\sigma$ tentative detection of the dipolar rotational kSZ effect (rkSZ) within the virial radii of our fast-rotating groups from SDSS with an average mass of $10^{14}\msunoh$ (corresponding to a null-hypothesis probability of 0.96\%). We estimate the averaged rotational velocity of the sample to be $\sim$\,$100-200\kms$, reaching a maximum value of $158.8\pm13.6 ~\kms$.%, with an additional 39.4\% systematic underestimate relative to the peculiar-velocity signal inferred from our simulations. 
Our results are consistent within the errors with predictions based on the ELUCID constrained realisation simulation, with the predicted amplitude of the rkSZ signal being slightly lower near the centre. We also identify a systematic bias when estimating rotational velocities using the observed redshifts of galaxies, but find it to be subdominant for our analysis. 
\end{abstract}


\section{Introduction}\label{sec::intro}
Galaxy clusters are thought to be the largest gravitationally bound structures in the Universe, formed through the hierarchical merging of smaller systems over cosmic time \citep[e.g.][]{1974_press_schechter,cluster_review_voit,Kravtsov_2012}. They are the composites of at least three major components. In decreasing order of mass fraction, these are dark matter, diffuse gas and galaxies (stellar component) -- the latter two being baryonic constituents. Understanding the baryon cycles around clusters and groups is a crucial step towards understanding galaxy formation, as well as its co-evolution with dark matter haloes in the Universe.  

The rotational motions of clusters and groups can provide unique constraints on the models for the formation of clusters, groups and the galaxies that they host, together with the transfer of angular momentum and the interplay between baryonic and dark matter. To what extent the rotational motions of dark matter, hot gas and galaxies are coupled in clusters is an open question, which matters for the understanding of the dynamical histories of baryons and dark matter in these systems. Recent analyses using hydrodynamical simulations suggested that there are connections between the angular momentum of the circumgalactic medium (CGM) and the star formation inside a galaxy \citep{Wang2022, Lu2022}. These authors predict a co-rotation of the CGM and the central stellar disk, and an anti-correlation between galaxy star-formation rates and orbital angular momenta of interacting galaxy pairs or groups \citep{Lu2022}. Some observational evidence for co-rotation between stellar discs and satellite galaxies was also reported \citep{Wang2024}. These results motivate further detailed studies of the rotational properties of material within groups and clusters.

In this context, it is of interest to note the prediction from general relativity that the rotation of a massive body should generate a frame-dragging effect called the gravitomagnetic effect. This means that rotating massive bodies such as groups and clusters can generate an additional gravitational lensing deflection, which supplements the usual lensing effect that arises purely from density \citep[e.g.][]{Ciufolini2003, Sereno2003, Sereno2005b,Sereno2005, Sereno2007}. This effect can potentially be detected using cross-correlations of lensing and other tracers of the rotating system \citep[e.g.][]{Schaefer:2005up, Barrera-Hinojosa2022}; see also
\cite{Andrianomena:2014sya,sagaWeakLensingInduced2015,Thomas:2015dfa,Adamek:2015eda,Cuesta-Lazaro:2018uyv,Crosta:2018var, Tang:2020com} for recent studies of the gravitomagnetic effect. It can serve as a test of general relativity, although the current observational sensitivity is many orders of magnitude away from the regime required for this.

Finally, non-zero gas rotational velocities will lead to incorrect cluster mass estimates, as these commonly assume hydrodynamic equilibrium. This in turn will affect the cosmological constraints
derive from cluster masses and number counts \citep[see e.g.][]{Rasia_2006,Rasia_2012,Matteo_2013,Gianfagna2021,Gianfagna2023}. 

To address some of the above questions, it is essential to build a clear understanding of the physical picture for the rotation of clusters and groups. In theory, clusters and groups are expected to acquire some spin as a result of tidal torques \citep[e.g.][]{Peebles_1969,White_1984}. In simulations, the angular momentum of dark matter haloes can be measured \citep[e.g.][]{Bett_2007,Maccio_2007,Zjupa_2017}, and the results imply that groups and clusters should have significant rotational support, at a few per cent to $\sim$\,10\% of the effect of randomly directed galaxy orbits \citep{Steinmetz1995, Cole1996}. 
However, it has been challenging to obtain direct observational evidence for the rotation of clusters. For the stellar component, a recent detection was reported using the redshift dipoles of satellite galaxies \citep{Tang2025}. For the hot gas, one possible way is to use the kinetic Sunyaev–Zeldovich effect (kSZ) \citep{SZ_1972, SZ_1980}. 

The kSZ effect arises from the Doppler shift imparted to CMB photons by the peculiar motion of free electrons. In the non-relativistic limit, this secondary CMB temperature fluctuation induced by kSZ is the integral of the electron momentum over the line of sight (LOS):
\begin{align}
    \Delta T_\mathrm{kSZ}(\hat{\bm{r}}) = - T_{\rm CMB}\, \sigma_\mathrm{T} \int n_\mathrm{e} \left( \frac{\bm{v}}{c} \cdot \hat{\bm{r}} \right) \mathrm{d}r,
    \label{eq:kSZ}
\end{align}
where ${\bm{v} \cdot \hat{\bm{r}} = v_\parallel}$ is the LOS component of the peculiar velocity of electrons, $\sigma_\mathrm{T}$ is the Thomson scattering cross-section, and $c$ is the speed of light. 

The sign of the kSZ effect is the opposite of the peculiar velocity -- a crucial feature to be exploited for its detection. For a gas halo moving towards/away from the observer, we expect a reduction/boost of the CMB temperature along the LOS. The kSZ effect has been used to measure the bulk motions of gas around galaxies, groups and clusters \citep[e.g.][]{Chen_etal_22, Li_etal_24, Hadzhiyska_etal_24, McCarthy2025, Lai2025, Hotinli2025, Hadzhiyska2025, Roper2025}. For a rotating halo of gas with its angular momentum perpendicular to the LOS, the gas on one side of the halo relative to its centre is moving towards the observer, and the gas on the other side is moving away. Therefore, a dipole of kSZ temperature fluctuation associated with the rotation of the gas is expected. This is the rotational kSZ effect (rkSZ) -- first pointed out by \citet{Cooray2002, Chluba2002}.

The rkSZ dipole is a unique signature associated with the spin of gas haloes. Its pattern is similar to that of the observed rotation of a spiral galaxy detected via Doppler redshift of spectrum lines, typically via HI or IFU observations \citep[e.g.][]{Rubin1980, Croom2012, Abdurrouf2022}. The rkSZ dipole is expected to be contributed by ionised gas (the inter-cluster medium, IGM and CGM), which may be prominent at the scales of clusters and groups. 

One major challenge for attempts to detect the rkSZ signal in this way is that we require prior knowledge of the direction of the gas halo's rotation axis, which informs us of the orientation of the temperature dipole on the sky. This is crucial if we are to use oriented stacking in order to beat down the noise. If we assume that the galaxies within a cluster are co-rotating with the IGM, we should expect to observe a dipole in the redshifts of member galaxies, which is aligned with the rkSZ dipole. This allows us to use the observed galaxy redshift dipole as a prior in looking for the rkSZ CMB dipole. We will adopt this approach below.

Another major observational challenge is that the expected rkSZ signal is very small. The temperature dipole associated with a rotating cluster is on the order of 10 $\mu$K or smaller \citep{Baldi2018, Altamura2023, Monllor‑Berbegal2024}. This is highly subdominant to the primordial CMB temperature fluctuations, but stacking a sample of rotating clusters can in principle, beat down the primordial CMB and other noise and allow a detection. Large and deep galaxy redshift surveys are needed in order to identify a suitable sample of clusters and their member galaxies. For this, we will use the group catalogue defined in the SDSS-DR7 galaxy sample \citep{Yang_halo_finder, Yang_etal_07}. 

Predictions of the rkSZ effect using simulations have been made in many studies: for example, \cite{ZorrilaMatilla_Haiman_21, Baldi2018, Altamura2023}. The first observational attempt was made by \cite{Baxter2019}, where a tentative detection of the rkSZ signal at the $\sim$\,$2\sigma$ level was found in the joint analyses of the 13 clusters selected from the SDSS-DR10 galaxy sample \citep{rotation_ref} and the CMB temperature map from \textit{Planck}. The low statistical significance of the detection was mainly due to the small number of clusters used for the stacking. 

In this study, we follow a similar approach to that taken by \cite{Baxter2019}, but extend the catalogue of galaxy clusters and groups found in the SDSS-DR7 galaxy sample\footnote{\url{https://gax.sjtu.edu.cn/data/Group.html}} based on the algorithm presented in  \cite{Yang_halo_finder, Yang_etal_07}. A particularly unique aspect of our study is that we employ a companion halo sample derived from a constrained realisation simulation of nearby large-scale structure -- the {\sc ELUCID} simulation \citep{ELUCID_16}. Properties of haloes found in the ELUCID simulation have been shown to match well with the groups identified from SDSS observations. This simulation provides us with guidance on how to optimise our observational selection and analyses, as well as how to make model predictions for interpreting the observed signal. 

The paper is organised as follows: in Section~\ref{sec::obs_data_and_sims}, we introduce the observational data and the constrained realisation simulation. Section~\ref{sec: selection} explains the selection of group catalogues, the individual matching with the constrained realisation simulation, and the estimate for the directions of angular momenta. Section~\ref{sec:model} presents our model of the observations using the constrained realisation simulation. 
Section~\ref{sec:observation} presents the main results of the observational analysis. We draw our conclusions in Section~\ref{sec:conclusion}.

\section{Observational data and simulations}\label{sec::obs_data_and_sims}
To measure the rkSZ signal, we need three pieces of information: \\
(1) a catalogue of groups with ionised gas; \\
(2) the projected angular momentum on the sky for each group; \\
(3) a CMB temperature map. \\
We will use the catalogue of galaxy clusters and groups derived from the SDSS-DR7 galaxy sample \citep{Yang_etal_07, Yang_etal_12} and the observed CMB temperature map from \textit{Planck}. We will employ a method used in \cite{rotation_ref} for determining the direction of angular momentum. In addition, we have the {\sc ELUCID} constrained realisation simulation dedicated to the SDSS-DR7 group sample at our disposal. We introduce these datasets in this section.

\subsection{ELUCID simulation and the SDSS group catalogue}
 The ELUCID simulation \citep[see][and references therein]{ELUCID_16} is a suite of high-resolution constrained realisation simulations. Specifically, the initial density field of this simulation is constrained by the present-day density field from the galaxy distribution in the Sloan Digital Sky Survey Data Release 7 \citep[SDSS DR7: see][]{Blanton_SDSS_DR7,Abazajian_SDSS_DR7}. The simulation evolves $3072^{3}$ dark matter particles, each with a mass of $3.09\times10^{8}~h^{-1}M_{\odot}$ in a cubic volume of $500~h^{-1}\rm Mpc$ on a side, starting from an initial redshift of $z_{\rm ini}=100$ under the WMAP5 cosmology \citep{WMAP5_Dunkley}.

To reconstruct the initial conditions, the Hamiltonian Markov Chain Monte Carlo method \citep[HMC, see][]{Wang_reconstruction_paper1,Wang_reconstruction_paper2} is employed to infer the initial density field in Fourier space from the observed present-day density distribution. 
The method adopts Particle-Mesh (PM) dynamics to compute the gravitational forces and therefore can accurately track the non-linear evolution of structure formation.
The present-day density field is constructed from a group-halo-domain density field derived from the SDSS DR7 galaxy catalogue. Galaxy groups are identified using an adaptive halo-based group finder, and group halo masses are then assigned via luminosity or stellar mass abundance matching \citep[see][]{Yang_halo_finder, Yang_etal_07}. The ELUCID simulation accurately reconstructs the present-day large-scale halo density field, albeit with uncertainties arising from redshift distortion corrections and limitations of the group-finding algorithm. 
For a more detailed discussion of the reconstruction method and a comparison between the reconstructed and true density fields, please refer to \cite{ELUCID_16}.

The one-to-one matching of the groups from the SDSS sample and the constrained realisation simulation allows us to use the simulated groups to tailor the observational selection, and make model predictions of the rotational kSZ signal. 

\subsection{CMB temperature map}
For the CMB map, we use the \textsc{smica-noSZ} map \citep{SMICA_noSZ_ref} \footnote{Map available at \url{https://irsa.ipac.caltech.edu/data/Planck/release_3/all-sky-maps/matrix_cmb.html}}. \textsc{smica} \citep[Spectral Matching Independent Component Analysis, see][]{SMICA_ref} is a component separation method that reconstructs CMB temperature and polarization maps in spherical harmonic space by linearly combining multi-frequency observations. The linear weights in \textsc{smica} are designed to preserve signals with the spectral dependence of the primary CMB anisotropies (i.e., blackbody spectrum), meaning that frequency-independent signals such as the kSZ effect are retained in the final map.

Meanwhile, the \textsc{smica} pipeline models and fits foreground components with different spectral behaviours, including Galactic dust, synchrotron emission, and the thermal SZ effect. As discussed in a later section, although our rkSZ estimator is weighted by the rotation velocities of galaxy groups and clusters, residual tSZ signals from these objects can still be non-negligible and bias the kSZ measurement. Therefore, we adopt the tSZ-free SMICA map release for our analysis instead of the original full-component \textsc{smica} map. The angular resolution of the map, determined by the parameter {\sc nside}, is 2048, which corresponds to 1.7 arcminutes. The FWHM of the beam for this map is $\theta_{\rm FWHM}$ = 5 arcminutes \citep{SMICA_noSZ_ref}.

\section{Selection of rotating groups}
\label{sec: selection}
Armed with the observational datasets and the ELUCID simulation, the major steps towards measuring the rkSZ signal are (1) identify the projected angular momentum on the sky for each group -- thereby the orientation of the rkSZ dipole on the CMB; and (2) orient and stack the CMB temperature maps using the projected angular momenta. The expected rkSZ signal is proportional to the projected gas density and rotational velocity. To maximise the signal-to-noise, we would like to select groups with large projected angular momentum i.e., massive and fast-rotating groups. We will exploit the constrained realisation simulation to optimise our selection.

\begin{figure*}
\includegraphics[width=\textwidth]{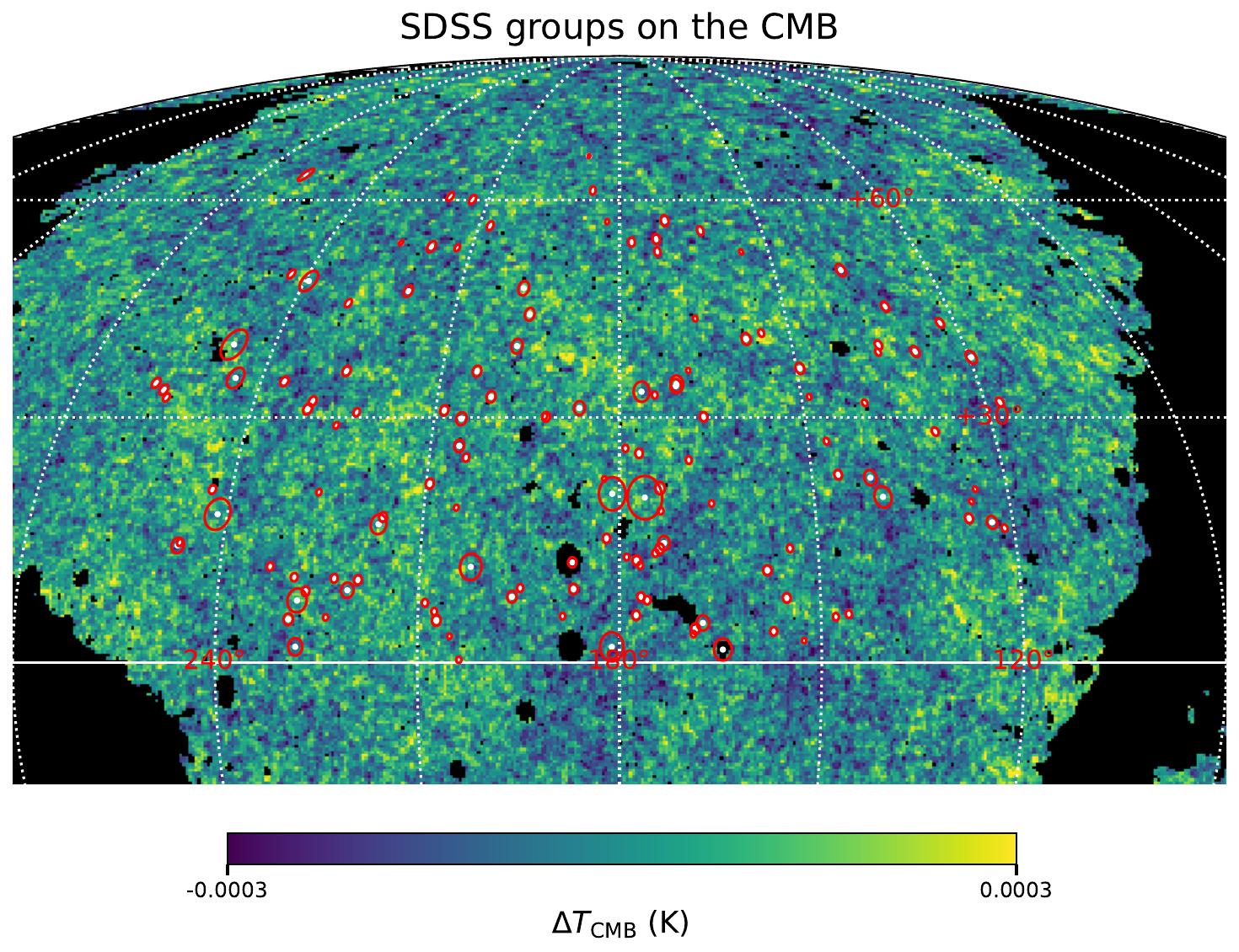}
   \caption{Mollweide projection of the sky positions of the selected 134 groups from the SDSS-DR7 parent sample overplotted on the \textit{Planck} CMB temperature map. Each white dot represents the group centre, and the associated red circle represents a $2\times r_{\rm vir}$ region around it. These are chosen to be fast-rotators with their rotation axes being approximately in the plane of the sky. They are individually matched to the haloes from the ELUCID constrained realisation simulation. The centres of the matched ELUCID haloes are indistinguishable from the white dots. The details of the algorithm for the selection of the groups, and their matching with the simulation are presented in Section~\ref{sec: selection}. The CMB temperature map with a mask applied is shown with the resolution parameter {\sc nside}$ ~=2048$.}
   \label{sky_positions}
\end{figure*}

\begin{figure*}
\includegraphics[width=\textwidth]{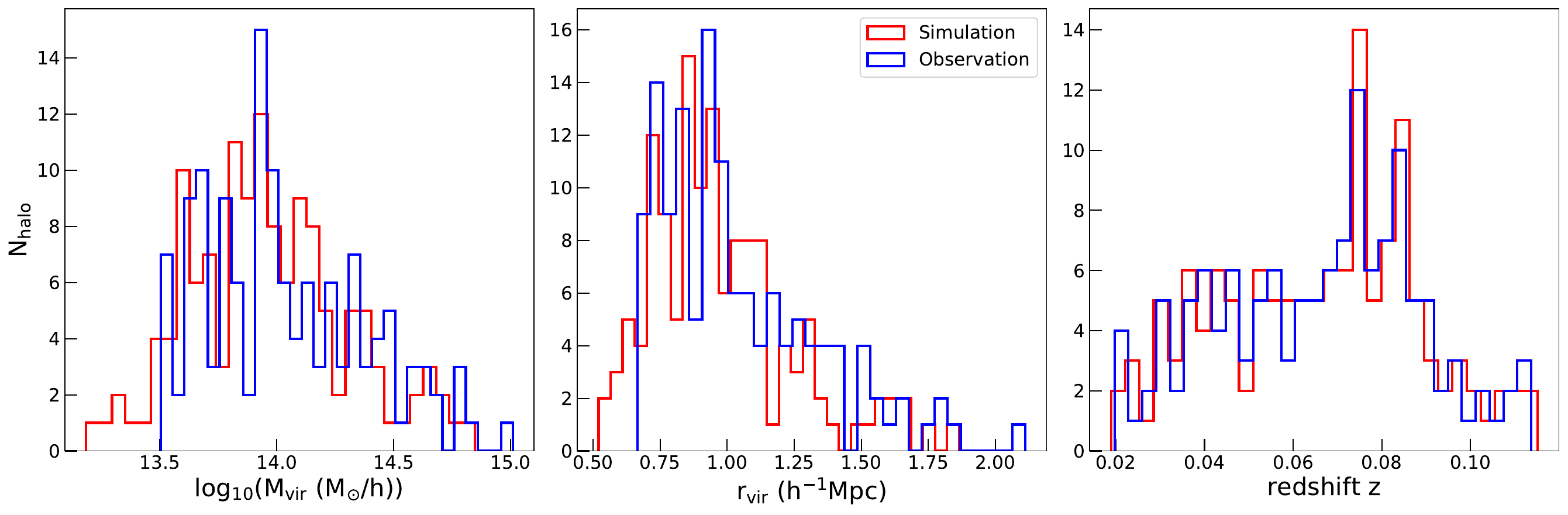}
   \caption{Histograms of the virial masses, radii and redshifts of the 134 selected SDSS-DR7 groups from the observation (blue) and their individually matched haloes from the ELUCID simulation (red). The mean redshift of the observed (simulated) sample is $\langle z \rangle \approx$ 0.066 (0.066), the mean virial radius $\langle r_{\rm vir} \rangle  \approx 1.04~(0.98)$ $\textrm{Mpc}/h$, and the mean mass is $\langle  \log_{10}{M_{\rm vir}~(h^{-1}{\rm M}_{\odot}) \rangle} \approx 14.19~(14.10)$. Part of the good agreement between the simulated and observed group masses is due to the abundance matching scheme applied to the simulation. See Section~\ref{sec: selection} for the details.}
   \label{sample_stats_summary}
\end{figure*}

\subsection{Matching observed groups to simulated haloes}
The ELUCID simulation provides us with a matching equatorial coordinate system, consistent with the observations of SDSS galaxies. This means that the simulation data are already placed in a lightcone, covering the same patch of the sky as the observational data. We can therefore select samples of haloes from the simulation that are individually matched with the observed SDSS groups. In practice, the locations and other properties of the simulated haloes may not be perfectly matched with the observed SDSS groups. The constrained realisation simulation is based on limited large-scale observational constraints, and was not designed to match smaller-scale structure. In general, the accuracy of the matching increases with the mass of a halo, as shown in Figure 9 of \cite{ELUCID_16}. Because of this, a mass threshold of $M_{\rm th}$ of $10^{13.5}\msunoh$ is applied to the observed list of groups in order to ensure the accuracy of the matching. Such a mass threshold is chosen to ensure that the majority of the matches remain physically meaningful and correlated with the observed large-scale structure, while still preserving a sufficiently large sample for the subsequent stacking analysis. No mass cut is applied to the simulated list of haloes. 

To perform the matching, starting from the most massive observed group, we search for all simulated haloes within a spatial distance $\Delta d_{\rm mx}$ and a mass difference satisfying $| \textrm{log} ~M_{\rm group}/M_{h}| \leq \Delta m_{\rm mx}$. If no simulated halo meets both criteria, the group is considered unmatched and disregarded. Otherwise, the halo with the closest mass to the group is selected as the best match. Once a match is found, both the group and the corresponding halo are removed from the lists, and the procedure is repeated for all the selected SDSS groups. We adopt $\Delta d_{\rm mx}$ and $\Delta m_{\rm mx}$ as $4 ~h^{-1}\rm Mpc$ and 0.5 dex for the spatial and mass matching thresholds, which allows us to individually match $\sim85\%$ of the observed groups with the simulated haloes. The matching between the observations and the groups forms the basis for the comparison between the constrained simulation and the observation. Despite the resultant matched results have some scatter, the groups have almost identical mass and redshift, on average. Later, we will use the matched haloes to make predictions for the rotation curves and the expected kSZ signal. 

% In total, 2472 groups matched with haloes are selected. 

However, not all of these matched groups are useful for the detection of the rkSZ signal. Only those with large transverse angular momenta are expected to contribute a large signal. For this, we will need to prune the sample further, starting by estimating the projected angular momentum. %In summary, there are 2966 groups above $M_{\rm th}$ in the full SDSS-DR7 group catalogue of 291659 systems, of which 2472 are matched with simulated haloes. From these matched systems, only 134 samples are identified by our fast-rotating group selection algorithm (as outlined in Section \ref{ssec::Fast-rotator_selection}), which is the final sample used for further analysis.

\subsection{Estimating the angular momentum}\label{ssec:: Estimating_the_angular_momentum}
For each object in the above matched catalogue, the orientation of its angular momentum i.e., the two-dimensional rotation angle, and the amplitude of the rotational velocity are determined using the method described in \citet{rotation_ref}. In summary, the method searches for a redshift dipole using the observed redshifts of satellite galaxies around each centre of a group. The axis of rotation is determined by the direction that maximises the amplitude of the redshift dipole, which is then treated as the proxy for the orientation of the projected angular momentum.

Using the observed galaxy and group samples in the SDSS area,  we first select all nearby galaxies around each group with stellar masses below $10^{10.5}\msunoh$, corresponding to host halo mass of approximately $10^{12}\msunoh$, located within a projected radius of $r_{\rm vir}$ and a redshift-space separation of $\Delta z=0.004$, corresponding to $\sim$\,$12\mpcoh$ in LOS distance separation around each group. The selection of satellites within transverse $r_{\rm vir}$ is informed by our test conducted with the ELUCID simulation in real space, where there are no complications from redshift-space distortions and interlopers. We find that rotation of matter is typically seen within $r_{\rm vir}$ in the test. The LOS redshift selection is to reduce the impact of potential interlopers. In the simulation, we similarly select all subhaloes with virial masses below $10^{12}\msunoh$ within the same projected radius and redshift separation around each halo centre as in the observed data. 

With the sample of subhaloes/galaxies selected in simulations/observations, we use the above method to select rotating systems. In the plane of the sky, the signature of a rotating group is identified using the tentative rotational motion of satellite galaxies, which would show a velocity pattern that changes its sign across the projected angular momentum axis, as a function of azimuthal angle $\theta_{\rm rot}$. For a given $\theta_{\rm rot}$, we have a measurement of the `rotation curve' $v(\theta | \theta_{\rm rot})$, where $\theta \in [0, 360^{\circ}]$.
In practice, we measure this rotation curve by binning in $\theta$, and determining the mean velocity and standard error in each bin. We assume that the galaxies are independent objects, so that the errors in each bin are uncorrelated.
We can then fit these measurements with a model sinusoidal rotation curve, yielding the goodness of the fit $\chi^{2}/{\rm d.o.f.}$ and the corresponding peak of the best-fitting rotation curve $v_{\rm rot}$. Here, $\chi^{2}$ is the chi-squared between the measured $v(\theta | \theta_{\rm rot})$ and the sinusoidal function; ${\rm d.o.f.}$ is the degrees of freedom i.e. the number $\theta$-bins;  We repeat this by sampling $\theta_{\rm rot} \in [0, 360^{\circ}]$ uniformly. The $\theta_{\rm rot}$ that maximises the velocity amplitude $v_{\rm rot}$ is defined as the rotation axis i.e., orientation of the angular momentum. The velocity difference on the two sides of this axis provides an estimate of the amplitude of rotational velocities projected on the sky.

However, there are subtleties arising due to redshift space distortions. In simulations, the redshift of each simulated satellite is given to second order as
\begin{equation}
\label{eq:redshift}
z_{\rm obs} = z_{\rm cos} + (1+z_{\rm cos})(v_{\rm pec}/c) = z_{\rm cos} + v_{\rm pec}/c +z_{\rm cos}v_{\rm pec}/c, 
\end{equation}
where $z_{\rm cos}$ is the cosmological redshift and $v_{\rm pec}$ is the peculiar velocity. While only the curl component of $v_{\rm pec}$ is responsible for producing the rkSZ signal, we only measure the total $z_{\rm obs}$. We have found in our simulation that the amplitudes of redshift dipoles selected using the observed redshifts of galaxies can differ from the true dipoles of peculiar velocities. This is because there are scenarios where the distribution of satellites is tilted along the line of sight, i.e., satellites on one side of the halo centre are intrinsically further away from the observer than those on the other side (having different cosmological redshifts $z_{\rm cos}$). In addition, because we are selecting all galaxies within a redshift slice of $\Delta z=0.004$ around each group, and in redshift space, there will also be contributions from nearby galaxies, whose distribution along the LOS can also be tilted. This can be interpreted by the algorithm as a rotational redshift dipole, leading to incorrect estimates of the actual rotational velocities in our matched sample, biased by up to 10's of per cent.

To make a fair comparison between the simulation and observations, we apply the method to the simulated/observed redshifts ($z_{\rm obs}$) of galaxies around each matched halo/group, respectively. That is, we perform the comparisons in redshift space. This accounts for the biased estimate of the rotational velocity from the simulated $z_{\rm obs}$, so that the simulated dipoles of peculiar velocities can be fairly compared to observations of the rkSZ. 

It is worth noting that the identified dipolar velocity pattern does not uniquely correspond to coherent rotational motion. Similar large-scale bipolar signals may also be generated by e.g. merger-driven bulk flows, or other non-equilibrium dynamical processes within the intragroup medium \citep[as discussed in e.g.][]{Goldstein2025}. To quantify this, we have tested the contribution of merger-driven bulk flows by separating the simulation sample according to the subhalo-to-central mass ratio, using a threshold of 10 to identify merger-like systems. We find no significant change in the rotation curve between the two subsamples, suggesting that merger-driven bulk flows do not significantly contribute to our measured signal. Still, while we use the term ``rotation'' for convenience and consistency with previous literature, the measured signal should more generally be interpreted as evidence for large-scale dipolar gas motion.

\subsection{Selecting fast-rotating groups}\label{ssec::Fast-rotator_selection}
To identify faster rotating groups, we impose the following selection criteria, as in \cite{rotation_ref}:
\begin{itemize}
    \item $\chi^{2}/{\rm d.o.f.}<1$ to ensure that the sinusoidal function is a good fit to the data;  
    \item $\chi^{2}_{\rm r}/{\rm d.o.f.}>1$, where $\chi^{2}_{\rm r}$ is the chi-squared between the measured $v(\theta | \theta_{\rm rot})$ and a randomised velocity curve. A large value of $\chi^{2}_{\rm r}$ insures that the observed rotational curve is well above the noise level;   
   \item and the ratio $\chi^{2}/\chi^{2}_{\rm r}<0.4$. Requiring a small value of this ratio enforces a stronger combination of the above two criteria. 
\end{itemize}
After applying these criteria, our final sample contains 134 matched fast-rotating systems from both the simulation and observations.  

Note that these selection criteria were set for the very massive systems in SDSS, and they may not be the optimal for our purpose of measuring the rotational kSZ signal. There may be room for improvement if we explore more broadly the selection criteria, but we may run into the risk of introducing a posterior bias. To avoid look-else-where effect, we choose to keep these criteria unchanged.

Fig.~\ref{sky_positions} presents the distribution of the 134 groups from the SDSS (white dots) on the sky. The sky positions of the groups and simulated haloes are well matched and are indistinguishable in the figure. 

Fig.~\ref{sample_stats_summary} shows the histograms of group mass, size, and redshift distributions of the sample. The mean mass, $\langle \textrm{log}_{10}(M_{\rm vir}~[h^{-1}{\rm M}_{\odot}]) \rangle$, of our simulated sample and SDSS groups is 14.10 and 14.19, respectively. The average size $\langle r_{\rm vir} \rangle$ is 0.98 and 1.04$\mpcoh$, and the mean redshift  $\langle z \rangle$ is 0.066 for both samples. The consistency in halo properties between the fast-rotating SDSS groups and simulated haloes shows the robustness of our matching procedure. Note that because of our selection on fast-rotating groups, the results we will obtain should not be interpreted as an ensemble average of all the groups above the mass-threshold. One should always bear in mind this selection effect. 

In summary, there are 2966 groups above $M_{\rm th}$ in the full SDSS-DR7 group catalogue of 291659 systems, of which 2472 are matched with simulated haloes. From these matched systems, 134 samples are identified by our fast-rotating group selection algorithm (as outlined in Section \ref{ssec::Fast-rotator_selection}), which is the final sample used in our analysis.

\section{Modelling the rotational kSZ signal}
\label{sec:model}

With the selected one-to-one matching of the groups between the true and simulated data, we will use the simulated haloes to make predictions for the expected rotational kSZ signal. This will be used to interpret our observational results presented in the next section. 

We first construct the rkSZ signal-only map using the dark matter particle field around the 134 selected fast rotators in our simulation. The amplitude of the rkSZ signal around central haloes is given by the integrated electron momentum along the line of sight:
\begin{equation}\label{eqn::rksz_eq_in_simulation}
   \frac{\Delta T_{\rm rkSZ}}{T_{\rm CMB}} = -\frac{\sigma_{T}}{c}\int^{z_{2}}_{z_{1}} \overline{n}_{e}(z) (1+\delta_{e})(\mathbf{v}_{e}-\mathbf{v}_{\rm cen})\cdot \hat{n} \frac{(1+z)cdz}{H(z)},
\end{equation}
where $\delta_{e}$ is the density contrast of free electrons, $\mathbf{v}_{e}-\mathbf{v}_{\rm cen}$ is the peculiar velocity of electrons relative to the halo centre, $\hat{n}$ is the unit vector of the line-of-sight direction. $\overline{n}_{e}(z)$ is the mean electron density of the Universe at $z$ given by
\begin{equation}\label{eqn::mean_ne_eq}
    \overline{n}_{e}(z) = \frac{\rho_{\rm b}(z)}{\mu_{e}m_{\rm p}},
\end{equation}
with $\rho_{\rm b}(z) = \rho_{\rm crit,0}\Omega_{b}(1+z)^{3}$ as the universal baryon density at the redshift of the halo taken from the best-fit constraint from {\it Planck} \citep{Planck2020_cosmo}; $\rho_{\rm crit,0}$ is the critical density in the universe at the present day. 
We take $\mu_{e} = 1.14$ as the mean molecular weight per free electron for a cosmic hydrogen abundance of $\chi = 0.76$, and $m_{\rm p}$ is the mass of the proton.

\begin{figure*}
    \begin{minipage}[b]{1.0\textwidth}
        \centering        \includegraphics[width=0.95\linewidth]{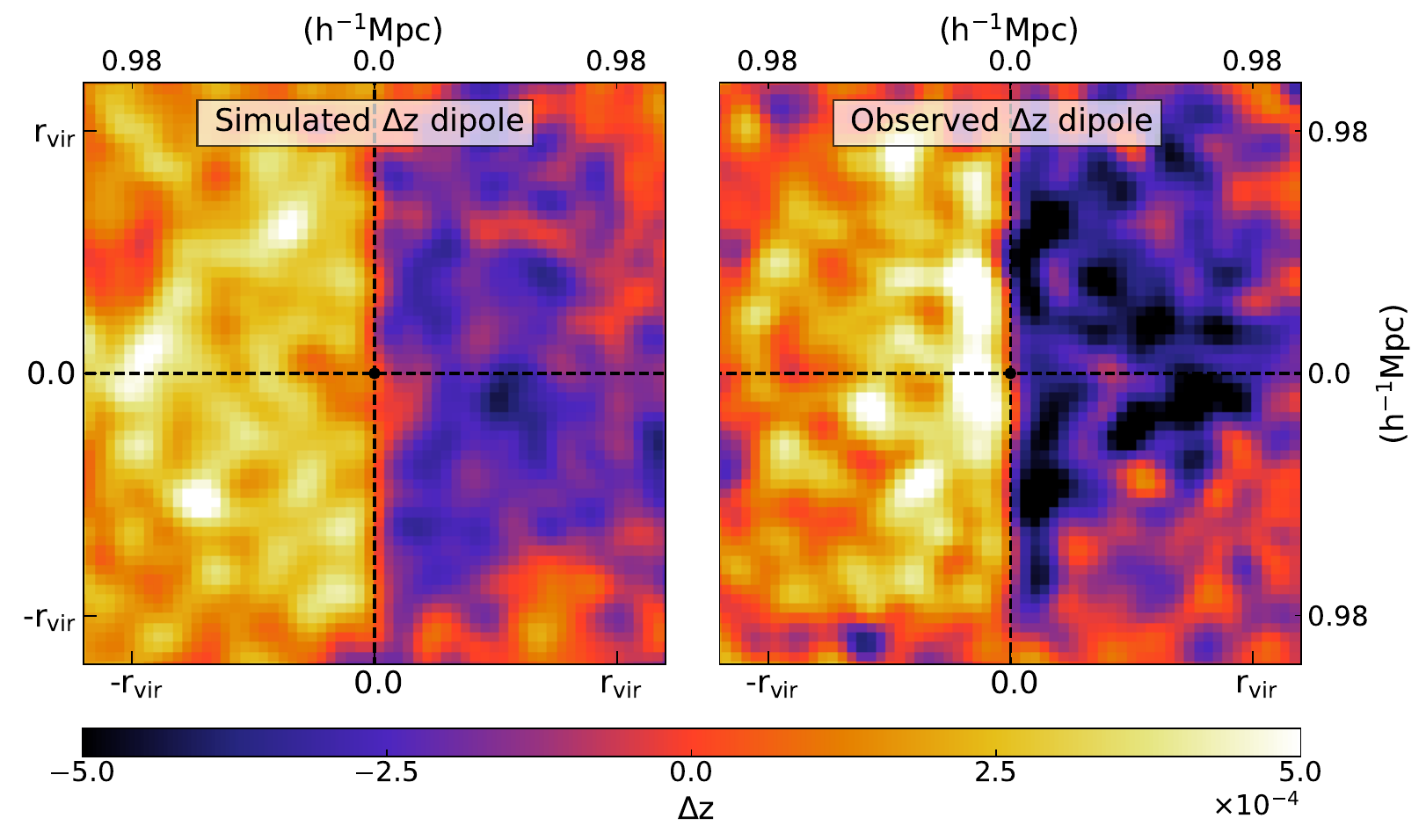}
        \includegraphics[width=0.95\linewidth]{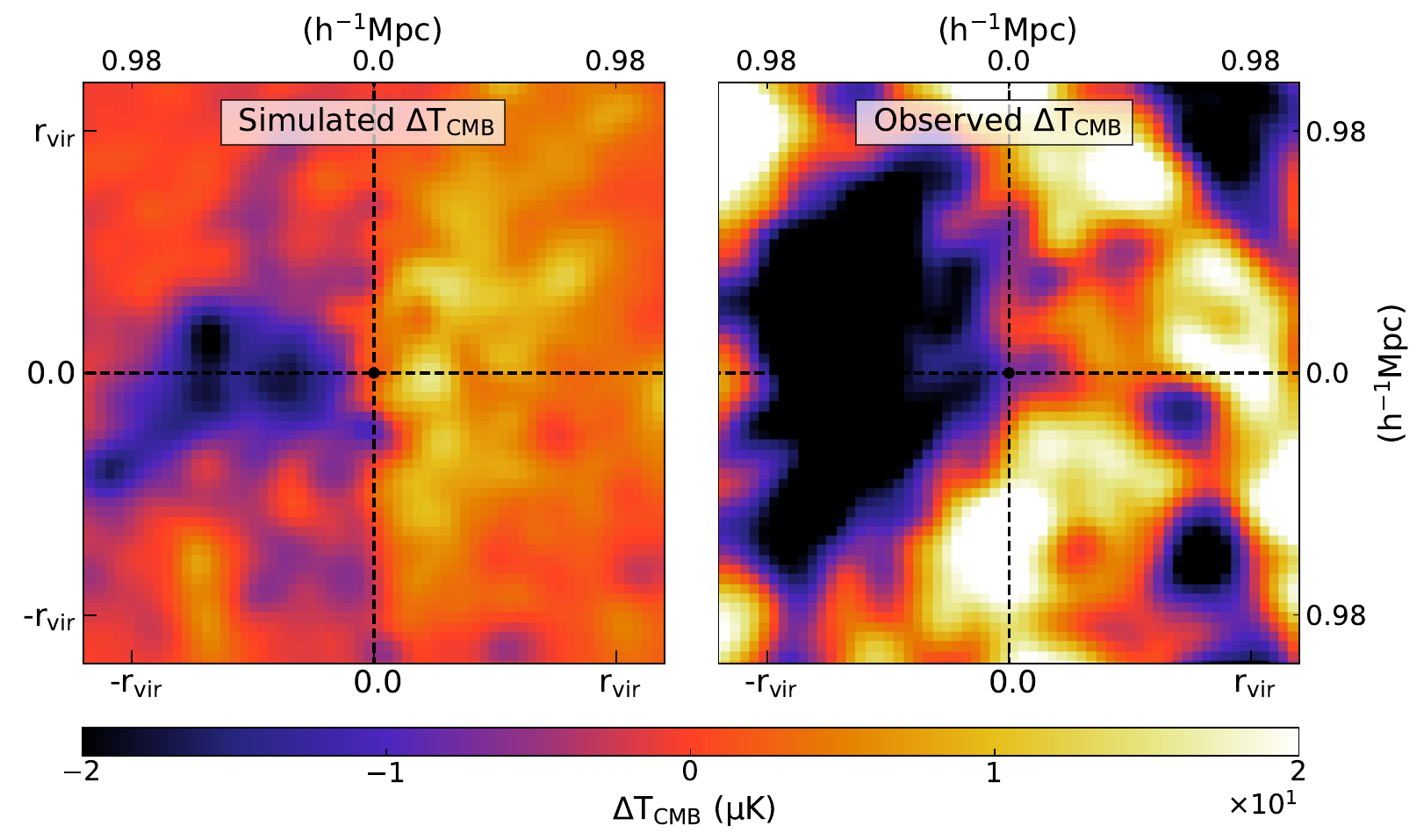}    
    \end{minipage}%
\vspace{+0.1cm}
\caption{\textit{Top-left:} The oriented stacked redshift map of simulated sub-haloes relative to the redshift of the centres of the 134 selected main haloes. \textit{Top-right}: the oriented stacked redshift map of SDSS satellite galaxies relative to the redshift of the 134 selected SDSS group centres. The virial masses and positions of group centres from the data are individually matched to the halo centres from the ELUCID simulation. All maps are viewed edge-on with the projected rotation axes aligned with the $y$-axis, pointing downwards in the direction of $-\vec y$.
\textit{Bottom:} the oriented stacked simulated and observed rkSZ temperature map of the 134 haloes (left) and groups (right). The rkSZ temperature fluctuation around each simulated halo is computed with Equation \ref{eqn::rksz_eq_in_simulation}. The signs of the temperature fluctuations at the bottom panels -- negative on the left and positive on the right -- are consistent with the signs of the redshift fluctuations -- redshift on the left and blueshift on the right at the top panels, if the redshift dipoles are responsible for the temperature dipoles via the rkSZ effect. For all maps,  a Gaussian filter of smoothing scale 0.06 $\mpcoh$ (1.5 pixels in units of $x, y$ grid) has been
applied to the image for illustration.}
\label{zredtot_sim_delta_T_four_panels}
\end{figure*}

Since the ELUCID simulation is a dark matter-only simulation, we assume that the matter and velocity distribution of electrons are unbiased against that of the dark matter, $\delta_{e} = \delta_{\rm dm}$ and $\mathbf{v}_{e} = \mathbf{v}_{\rm dm}$. To isolate the signal induced by the rotational field of the central, we only include dark matter particles within $\pm 2r_{\rm vir}$ around each object \footnote{We have tested the dependence of the rotational signal on the line-of-sight integration range and find that it approximately converges by $2r_{\rm vir}$, with no significant change when the integration range is extended to $4r_{\rm vir}$.}. With the positions and velocities of particles in the simulation, we assign the momentum to a $100^{3}$ mesh around each halo centre. We then integrate along each LOS to project the momentum field onto the plane of the sky. 
This yields a predicted rkSZ map associated with each selected group. We can then rotate these maps to align the projected angular momentum vectors for each group; this yields a prediction for the average rkSZ signal of the selected groups, presented in the bottom-left panel of Fig.~\ref{zredtot_sim_delta_T_four_panels}. 
    
\begin{figure*}
    \begin{minipage}[b]{1.0\textwidth}
        \centering        \includegraphics[width=0.487\linewidth]{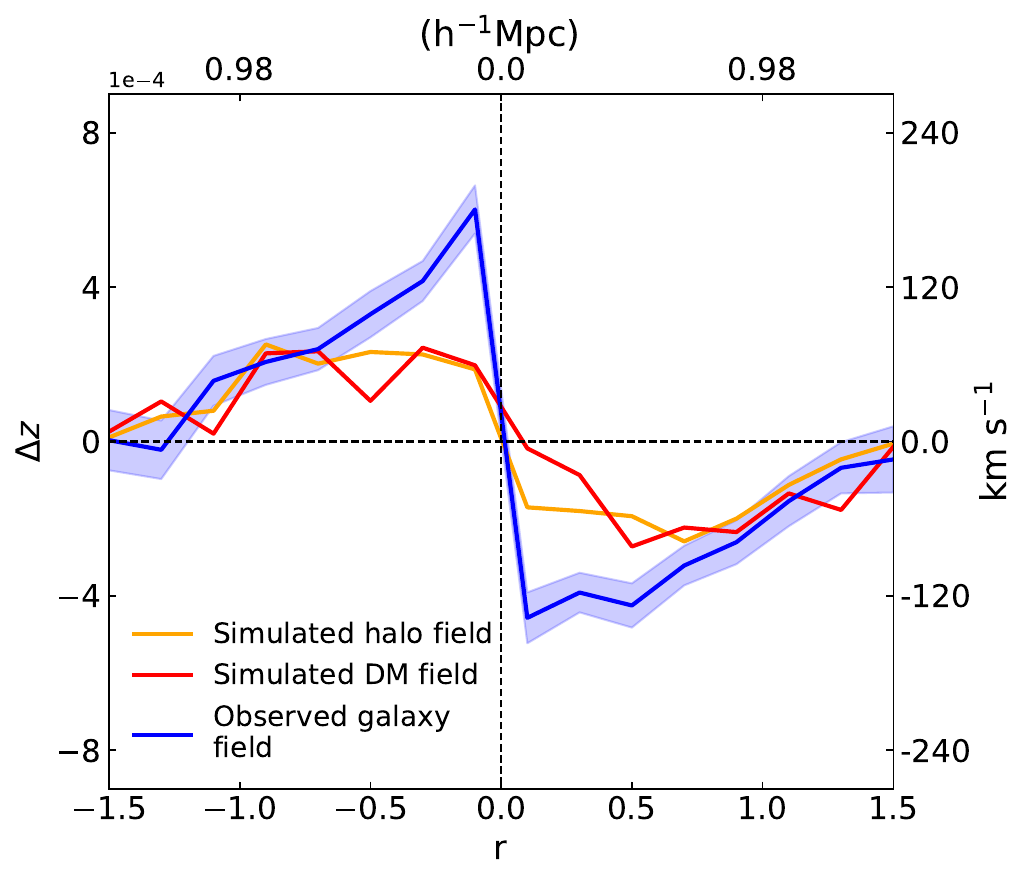}
        \hspace{0.2 cm}
        \includegraphics[width=0.487\linewidth]{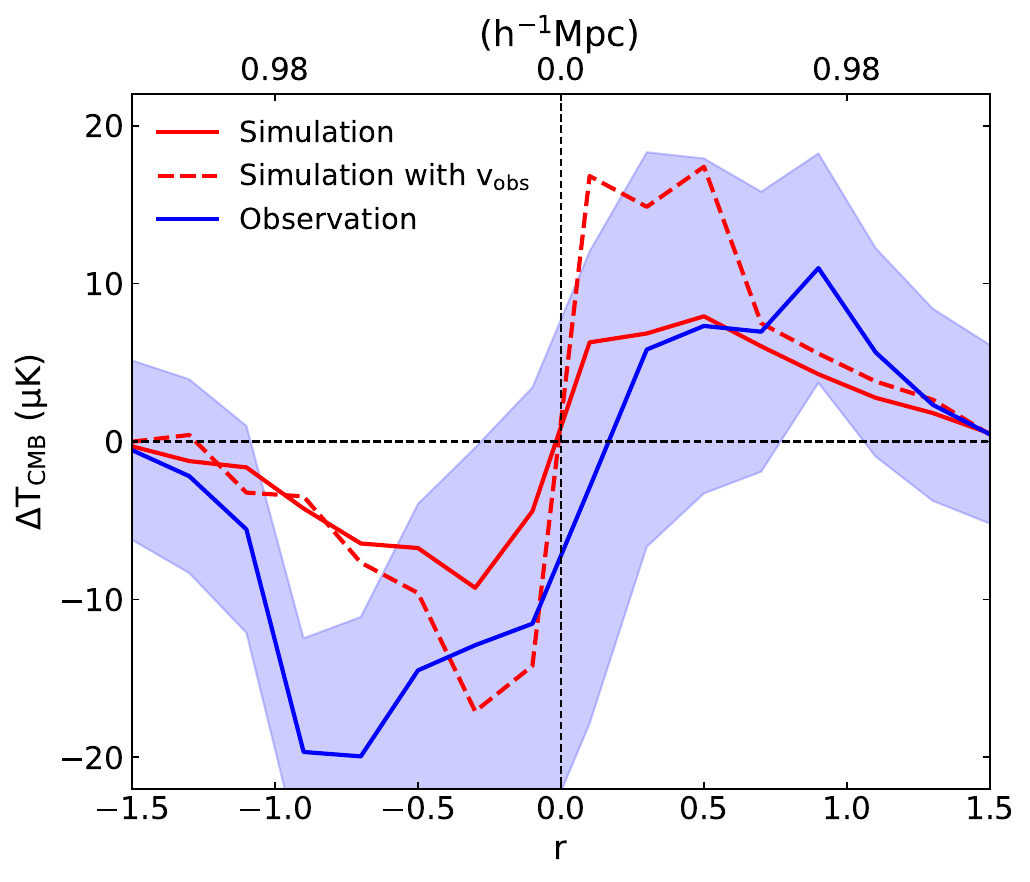}    
    \end{minipage}%
\vspace{+0.1cm}
\caption{\textit{Left:} The horizontal redshift profiles extracted from the top panels of Fig.~\ref{zredtot_sim_delta_T_four_panels}. For each radial bin, we draw an annulus from the centre of the stack and take the difference between the average redshift within the semi-annulus on the left versus the one on the right. Orange -- redshifts traced by subhaloes in the ELUCID simulation; Red --  redshifts traced by the dark matter field in the simulation; Blue -- the observed redshifts of SDSS satellite galaxies. The shaded regions represent the error on the mean of the 134 groups. \textit{Right:} The horizontal temperature profiles extracted from the stacked simulated (red-solid line) and observed (blue) $\Delta \rm T$ map from the bottom panel of Fig.~\ref{zredtot_sim_delta_T_four_panels}, respectively. Shaded regions represent the 1-$\sigma$ errors estimated with 10\,000 Gaussian realisations of the CMB temperature map. The red-dashed line is the prediction using the simulated dark matter field, but assuming that their rotation follows the rotation curve measured from the observed redshift dipole of the galaxy field -- the blue curve on the left-hand panel.}
\label{horizontal_profiles_zredtot_kSZ}
\end{figure*}

\section{detection of the rotational kSZ signal}
\label{sec:observation}

With 134 selected groups and haloes from the observed data and correspondingly from the ELUCID simulation, along with their estimated angular momenta, we can now perform oriented stacking of the redshift map from SDSS galaxies and \textsc{smica-noSZ} CMB temperature map from {\it Planck} in each sample.

\subsection{Oriented stacking}
\label{sec:stack}
We translate each temperature map such that the centre of the halo/group is at the origin of the 2D Cartesian coordinates. We then rotate each temperature map according to the estimated direction of the projected angular momentum, such that the $y$-axis in the plane aligns with the rotation axis. We choose the convention that the direction of the angular momentum points downwards, along $-\vec y$. We normalised each map by the virial radius $r_{\rm vir}$ of each halo in the simulation (and group from observation). The stacked map of rkSZ is weighted by an estimate of the angular momentum as follows:
\begin{equation}\label{eqn::rksz_stacked_eqn}
    \langle \Delta T_{\rm CMB }({\vec \theta}) \rangle = \frac{\Sigma^{N_{g}}_{1}N_{\rm sat,i}v_{\rm rot, i}\Delta T_{\rm CMB}(\vec \theta)M_{\rm T, CMB}(\vec \theta)}{\Sigma^{N_{g}}_{1}N_{\rm sat,i}v_{\rm rot, i}M_{\rm T, CMB}(\vec \theta)},
\end{equation}
where $N_{g}$ is the number of groups; $N_{\rm sat, i}$ is the total number of satellites around the $i$th group. $v_{\rm rot, i}$ is the rotation velocity (see Section \ref{ssec:: Estimating_the_angular_momentum} for its definition). We adopt such weighting $N_{\rm sat}$ and $v_{\rm rot}$ as an observationally motivated proxy for the expected rkSZ contribution from each group. Here, $N_{\rm sat}$ provides a simple observational tracer of halo richness and is expected to correlate with halo and gas mass, while $v_{\rm rot}$ measures the amplitude of the inferred dipolar velocity field. This weighting is therefore intended to approximate the dependence of the rkSZ signal on gas momentum,
% The product of $N_{\rm sat, i}$ and $v_{\rm rot,i}$ gives an estimate of the angular momentum, 
treated as weights to boost the contribution of the massive and fast-rotating haloes/groups. $M_{\rm T, CMB} (\vec \theta)$ is the mask of the {\it Planck} CMB temperature map. We use the \texttt{COM-Mask-CMB-common-Mask-Int-2048-R3.00.fits} mask map from \textit{Planck}. In the simulation, there is no mask and so $M_{\rm T, CMB} (\vec \theta)=1.0$. 

To suppress the large-scale variance from the primordial CMB temperature fluctuations, we apply a spatial filter to the CMB map before stacking. The angular scale of the virial radii of the groups ranges from $\sim$10 to $\sim$100 arcminutes. In this regard, we filter out angular scales larger than 110 arcminutes (corresponding to $\ell \sim 100$). Around each group centre, we extract a square patch of $4^{\circ}\times4^{\circ}$ from the sky map. The size of the patch is sufficient to encompass the maximum angular extent of the projected virial radius in the largest group, which corresponds to $\sim$\,$1.5^{\circ}$. Each patch is normalised by the angular scale of $r_{\rm vir}$ and then rotated according to its estimated direction of the projected angular momentum.

We apply the same stacking procedure to the simulated and observed redshift maps created using the observed/simulated redshift of galaxies/haloes to obtain the stacked redshift maps:
\begin{equation}\label{eqn::z_stacked_eqn}
    \langle z_{\rm rot} (\vec \theta) \rangle = \frac{\Sigma^{N_{g}}_{1}n_{\rm sat,i}(\vec \theta)v_{\rm rot, i} z_{\rm obs}(\vec \theta)}{\Sigma^{N_{g}}_{1}n_{\rm sat,i}(\vec \theta)v_{\rm rot, i}},
\end{equation}
where $n_{\rm sat,i} (\vec \theta)$ is number of satellites of the $i$th group in each 2D pixel, and $N_{\rm sat,i}=\int n_{\rm sat,i} (\vec \theta) d \vec \theta $. This ensures that the weights entering the stacked redshift map are equivalent to those for the stacked temperature map. 

\subsection{Redshift and temperature dipoles}
We apply the above oriented stacking to the redshift maps and temperature maps in both the simulation and observations. The results are presented in Figs.~\ref{zredtot_sim_delta_T_four_panels}. 

We can see strong dipolar patterns for all the redshift maps and temperature maps in Fig.~\ref{zredtot_sim_delta_T_four_panels}. This indicates the success of our estimates for the directions of the projected angular momentum. With the angular momentum pointing downwards along the $-\vec y$ direction, the satellites are expected to move away from the observer on the left (redshift) and towards the observer on the right (blueshift). This is seen in both simulation (top-left panel) and observation (top-right panel). The observed redshift maps are strikingly similar to the simulated version in their coherence within the virial radius. This indicates the similarity of the observed and simulated samples in phase space, although it is noticeable that the observed redshift dipole has a higher amplitude near the centre. 

We extract the amplitudes of the redshift/blueshift by taking the average values from the 2D maps within semi-annuli of different radii. The results are presented on the left-hand panel of Fig.~\ref{horizontal_profiles_zredtot_kSZ}. We can see that the observed redshift dipoles are of the order of $10^{-4}$, corresponding to 100-200$\kms$.
%with a maximum value of $158.8\pm13.6~\kms$.
At $r\simeq0.1r_{\rm vir}$, the mean of the absolute values of the positive and negative peaks gives a characteristic rotational velocity of $158.8\pm13.6~\kms$. It is worth noting that the quoted uncertainty here is the statistical uncertainty on the mean galaxy-redshift dipole of the 134 groups and does not include the uncertainty associated with the rkSZ detection, as discussed below.
The amplitude of the observed redshift dipole appears to be larger than the simulated version at small radii (0.2-0.5 $r_{\rm vir}$), but they agree well at larger radii, possibly due to the different properties of sub-haloes versus satellite galaxies near their respective centres. We will discuss this below.

We find good correspondence between the predicted oriented stacked redshift map and the rkSZ map in both simulation and observation, as shown in Fig.~\ref{zredtot_sim_delta_T_four_panels}. This indicates that there is a close connection between the rotation of satellite galaxies and the rotation of gas in the observed data (or between the rotation of sub-haloes and the rotation of dark matter in the simulation). The amplitude of the observed rkSZ signal peaks at approximately $10-20~\mu$K (right-hand panel of Fig.~\ref{horizontal_profiles_zredtot_kSZ}). It appears to be slightly larger than predicted from our simulation (red-solid line in the plot), but they are consistent within the errors. 

Given that the observed amplitude of the redshift dipole is greater than the prediction of our simulation, we try the experiment of making another prediction by applying the {\it observed} redshift dipole as the rotation curve to our simulated matter field. This is shown in the red-dash line on the right-hand panel of Fig.~\ref{horizontal_profiles_zredtot_kSZ}. As anticipated, we find the amplitude of the predicted temperature dipole increases at small radii, but they are still consistent within the errors with the observed version. This test gives us an estimate for the upper bound of the rotational velocity of the observed groups, which peaks at approximately 200$\kms$. 

Another interesting finding arises from comparing the observational results between left and right plots in Fig.~\ref{horizontal_profiles_zredtot_kSZ} -- the galaxy redshift profile peaks at much smaller radii than the rkSZ profile. Though both galaxy and gas co-rotate with their host halo, it seems that they behave differently with respect to the dominant dark matter component, which could be due to their collisionless vs hydrodynamic natures. One might expect that the early baryonic collapse of the halo may enhance the gas rotation of the inner CGM, which remains imprinted even as the halo grows much larger.  However, we can not verify this with the ELUCID simulation, because it only models dark matter.

We emphasise again that the stacked map of the total redshift is not identical to the peculiar velocity map, and only the latter is responsible for the rkSZ signal (see the discussion in Section~\ref{ssec:: Estimating_the_angular_momentum}).
% However, in practice, the difference between the redshift dipole and the velocity dipole for our sample is found to be at the order of tens of per cent. Such a difference is subdominant to the observational errors, so we do not intend to address this issue in the present paper. 
In practice, we find from our simulations that the redshift dipole can underestimates the peculiar-velocity dipole by approximately 35-45\% over the radial range considered. This is a substantial fraction of the signal. This systematic is alarming for estimating the rotational velocity using the observed redshift dipole alone. However, in the kSZ analysis, only peculiar velocities will give rise to the kSZ signal when we correlate the observed redshift dipole with the CMB, there is little concern that this may cause a false detection. In addition, since we propagate this systematics in our predictions using the constrained simulation, similar systematics exist in both the prediction and the observation, and so the comparison between them is valid. 
%, with a $\sim$40\% underestimate at the peak. %We therefore regard this difference as a systematic uncertainty associated with using the total redshift to infer the rotational velocity. We do not attempt to correct the measured velocity directly, since the magnitude of this bias may depend on the properties of the halo sample and the observational reconstruction.
The resemblance between the simulated and observed dipolar signal, especially with the one-to-one matching between the haloes and the groups, strongly indicates that the observed temperature dipole has a rotational kSZ origin. We will now assess the statistical significance of the signal.

\begin{figure}
\includegraphics[width=\columnwidth]{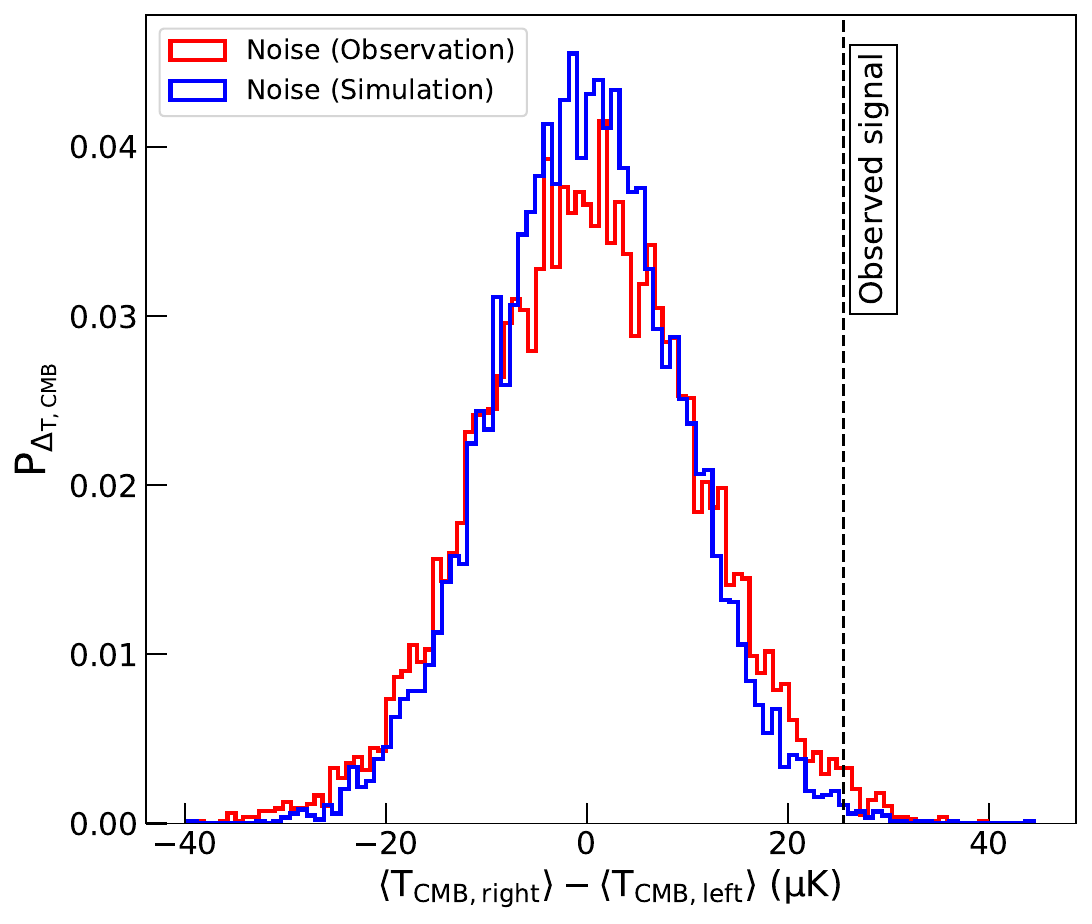}
   \caption{PDFs of the temperature difference $\Delta T_{\rm CMB}$ between the left and the right semi-circular regions of the stacked CMB noise maps. Blue is the result obtained from 10\,000 simulated CMB maps, including instrumental noise. Red is the result from the observed \textit{Planck} \textsc{smica-noSZ} CMB temperature map with randomly offset group centres such that there is no rkSZ signal. The vertical dashed line indicates the same detection for the observed signal from \textit{Planck}, at $\Delta T=25.5\,\mu$K. The standard deviation is $ 10.7\,\mu$K with the noise estimated from observations (or $ 9.3\,\mu$K with the noise estimated from simulations). The corresponding probability of this signal arising due to statistical fluctuations is 0.96\% (or 0.35\% from simulated noise maps), at the $\sim$\,$2.3\sigma$ (or $\sim$\,$2.7\sigma$) level.}
   \label{delta_kSZ_hist}
\end{figure}

\subsection{Statistical significance of the rkSZ signal}
To evaluate the statistical significance of our detection, we employ two different approaches to generate noise maps for cross-validation. We will then repeat the same analysis with the noise maps to estimate the corresponding noise levels. We evaluate the statistical significance of the signal by comparing the observational signal with the estimated noise levels.

\subsubsection{Constructing noise maps}
First, we build the noise matrix directly from the observational data. Using the groups and the CMB map from observations, we make random offsets between the group coordinates and the CMB coordinates via rotation, and perform the same oriented stacking for each realisation. We repeat this procedure 10\,000 times. This allows us to quantify the noise level arising from the observed CMB temperature fluctuations and instrumental noise. A caveat is that the large number of random offset realisations may not be truly independent from each other due to the finite size of the sky area. 

In the second method, we assume that the noise is dominated by the primordial CMB temperature fluctuation and the instrumental noise of {\it Planck}. This is justified by the fact that the kSZ signal and other CMB secondary contributions are subdominant. With this assumption, we can build the noise matrix contributed by the primordial CMB temperature fluctuation and the instrumental noise. We generate a sample of Gaussian realisations of the primordial CMB maps with noise. The primordial CMB angular power spectrum is generated using CAMB \citep{CAMB_ref}, while the instrumental noise power spectrum is modelled by:
\begin{equation}\label{eqn::instru_noise_level}
    C^{\rm noise}_{\ell} = \left(\frac{\Delta_{T}}{T_{\rm CMB}}\right)^{2}
\end{equation}
where $\Delta_{T}=3.1\,\mu$K-arcmin corresponds to the noise level for the \textit{Planck} temperature map \citep[e.g.][]{Knox:1995dq}. We then perform the same oriented stacking with each of the synthetic CMB+noise maps as we did with the observational map. The sky locations of the groups and their projected angular momenta are kept the same as before. This procedure -- from power spectrum to map generation and stacking -- is repeated 10,000 times to build a robust statistical baseline. 

These maps allow us to estimate the errors presented in the right-hand panel of Fig.~\ref{horizontal_profiles_zredtot_kSZ}, and to calculate the statistical significance of the observed kSZ dipole signal.

\subsubsection{Quantifying the signal-to-noise}
To quantify the statistical significance of the stacked signal, we reduce the observed dipole into a signal value of temperature. We take the difference of the mean temperature between the left and the right of the stacked map i.e., $\langle \Delta T_{\rm CMB}\rangle = \langle T_{\rm CMB, right}\rangle-\langle T_{\rm CMB, left}\rangle$, where $\langle T_{\rm CMB, right(left)}\rangle$ is the mean temperature on the right (left) semi-circular disk relative to the stacking centre, with a radius equal to the stacked $r_{\rm vir}$. The measured amplitudes for observation and simulation (bottom panel of Fig.~\ref{zredtot_sim_delta_T_four_panels}) are $\smash{\Delta T^{\rm Obs}_{\rm CMB}} = 25.5\,\mu$K and $\smash{\Delta T^{\rm Sim}_{\rm CMB}} = 12.8\,\mu$K, respectively. Repeating the same procedure using the noise-only maps yields the PDFs presented in Fig.~\ref{delta_kSZ_hist}. We estimate from the  10\,000 noise realisations that the standard deviation is $ 10.7\,\mu$K ($ 9.3\,\mu$K from Gaussian simulations). This corresponds to a 2.3$\sigma$ (2.7$\sigma$) tentative detection. So, there is a 0.96\% (0.35\%) chance that our detection is due to statistical fluctuations.

The two different ways of estimating the noise yield results that are consistent with each other within $10\%$. The errors estimated from the Gaussian realisations are slightly smaller, as shown by the histograms in Fig.~\ref{delta_kSZ_hist}. This is perhaps due to the extra noise contributed by other CMB secondary effects and the sample variance of the rkSZ signal itself, which is present in the observed CMB noise, but not in the Gaussian simulations. The good agreement between the two estimations of the noise confirms the robustness of our estimation of the statistical significance of the rkSZ dipole. With the $\sim$\,$10\%$ uncertainty on our estimated errors in mind, we report our result to be 2.3$\sigma$, which is a conservative choice.

\section{Conclusions and discussions}
\label{sec:conclusion}
The rotational kSZ effect from clusters and groups of galaxies is expected to generate a coherent temperature dipole imprinted on the CMB. In this paper, we conduct a search for such a signal via oriented stacking of the CMB temperature map using groups defined in the SDSS-DR7 galaxy sample \citep{Yang_halo_finder, Yang_etal_07}. A crucial element of this analysis is the knowledge of the direction of the projected angular momentum for each group, which is expected to be aligned with the dipole of rkSZ. We estimate the directions of angular momenta of the groups using the observed redshifts of satellite galaxies -- the redshift dipole, following the method presented in \cite{rotation_ref}. We detect a redshift dipole with a high significance for our selected 134 SDSS groups, and a rkSZ dipole associated with it at the level of $\sim$\,$2.3\sigma$. The tentative rkSZ signal reaches its maximum amplitude at $\sim$\,$25\,\mu$K. 

The statistical significance of the measurement is not high. To investigate which population of the selected 134 groups may be the dominant contribution to the tentative signal, we repeat our measurement with different weighting schemes for our stacking, and re-assess the statistical significance. When we adopt the group-mass as the weight, we find the significance drops to $1.4\sigma$, lower than the results from the fiducial ($N_{\rm sat}~v_{\rm rot}$) weighting scheme. This suggests that some massive groups may not be fast-rotating in their gas components, which, once up-weighted, can dilute the statistical significance. Indeed, when we adopt equal-weights for all groups, we recover a $\sim 2.5\sigma$ measurement, which is comparable to our fiducial weighting scheme. This indicates that both the mass the rotational velocity are important for a group to contribute to the rkSZ signal, as expected. This also indicates that our measurement is perhaps not completely dominated by the rare systems with large values of $N_{\rm sat}~v_{\rm rot}$'s.

%which is reassuring since the detection is not particularly dependent on the fiducial weighting, using equal weights actually gives a comparable significance, although indeed, based on the $N_{\rm mem}$ test, rich systems contribute more signal, which is physically expected. 

To model the signal, we simulate the rkSZ signal assuming a universal baryon fraction and that free electrons are unbiased tracers of dark matter. We then perform individual matching between the groups from the observed data and the haloes from the ELUCID constrained realisation simulation. We apply the same algorithm for selecting groups and haloes to both datasets, estimating the direction of angular momenta, and perform the same oriented stacking for the redshift dipole and temperature dipole. This ensures that selection effects are accounted for as far as we could when using the simulation to make model predictions of the redshift dipole and rkSZ dipole. The resulting simulated redshift dipole and the temperature dipole appear to be broadly consistent with the observed version, given the relatively large statistical error of the data. The comparison between the simulated and the observed rkSZ signal allows us to estimate the rotational velocity of our sample, which is approximately 100-200$\kms$ with a maximum of $158.8\pm13.6~\kms$. 

We do find the observed redshift dipole to be stronger near the SDSS group centres than in the simulated version. This is also consistent with the fact that the predicted rkSZ signal appears slightly lower than the observed one. This seems to suggest that there is a similar level of strong coupling between the rotation of sub-haloes and the rotation of dark matter in simulations, versus the rotation of satellite galaxies and the rotation of hot gas in observations. So, the slight discrepancy between the simulated and observed rkSZ signal probably reflects an imperfect match between the simulated haloes and the observed groups, despite the fact that the simulated one is constrained by observational results. 

One possible source of discrepancies between our observations and the constrained simulation is the effect of tidal stripping. In N-body simulations, which have no baryons, subhaloes are more likely to suffer from tidal stripping near the centre of their main halo than at the outskirts. Once the subhalo masses fall below the mass resolution of the simulation, they will dissolve \citep[e.g.][]{Gao2004, Giocoli2008,Han2016, Errani2020}. This can lead to a relative suppression of the number density of subhaloes near the centre. But real satellite galaxies have deeper potential wells, due to their compact stellar component, and are more likely to survive near the group centre. Indeed, we have checked explicitly that the stacked galaxy number density profile in our observations is steeper than the subhalo number density profiles in our simulation. Also, it is known from previous simulations that the direction of angular momentum can change significantly between the centre and the outskirts of a halo \citep[e.g.][]{Bett2010}.
Combining these two effects, it is not surprising that the measured rotation curves and the direction of angular momentum differ between simulation and observation. The observed version will receive more contributions from satellites near the centre, and thus the measured directions of angular momenta are more aligned with the central value. On the other hand, the simulated directions of angular momenta are better aligned with the values in the outer parts of the halo. This leads to a relative suppression of the amplitude of the simulated rotation curve at small radii.

Admittedly, the ELUCID simulation would only ever have been expected to give approximate predictions of the locations and masses of the groups, but even in this 2D parameter space, the discrepancies in these quantities can be of order unity for some groups. It would be even more difficult to pin down the internal structure and motions of sub-haloes of an individual halo, which matters for the predicted rkSZ signal. Above all, the simplifying assumption that the gas distribution and velocities follow that of dark matter is likely to contribute to some biases and uncertainty. Given the relatively low significance of our detection, we do not attempt to match the simulated sub-haloes with the observed satellite population, which may be what is eventually needed to increase the accuracy of the prediction.

%It is worth noting that we made a simplified assumption when predicting the rkSZ signal from the DM-only ELUCID simulation, namely that the dark matter traces the gas ($\delta_{e} = \delta_{\rm dm}$ and $\mathbf{v}_{e} = \mathbf{v}_{\rm dm}$ in Equation \ref{eqn::rksz_eq_in_simulation}). As discussed in \citet{Altamura2023}, the spin direction of gas and dark matter may not align in their hydrodynamical simulation. The resulting rkSZ prediction can differ by a factor of $\sim$1.5 to 2, depending on which spin direction was adopted. Our DM-only rkSZ prediction should suffer from such a bias, if it were proven to be true. %A more realistic treatment of the gas dynamics with hydrodynamical simulations will be explored in future work.

Note also that the stacked profile as presented in Fig.~\ref{horizontal_profiles_zredtot_kSZ} could be affected by several sources of intrinsic scatter. The halo virial radii are inferred from abundance-matched halo masses and therefore carry statistical uncertainties, while the adopted group centres may not coincide exactly with the true centres of the gas or dark matter distributions. Both effects tend to broaden the stacked profile and suppress its amplitude. Therefore, the rkSZ 1D profile should be interpreted as the average profile of the sample rather than a universal profile for individual systems. Consequently, the radius at which the stacked profile reaches its largest amplitude should not be regarded as a precisely determined physical scale.

We further caution that the interpretations of our results are based on a tentative detection with a $2.3\sigma$ significance, but the prospect of increasing the size of the sample is very bright, with more spectroscopic redshift data being collected by surveys such as DESI \citep{DESI_DR1}, and high fidelity CMB data from the Simons Observatory \citep{Abitbol2025}.   

% Considering the significant difference in the selection of groups in our observational analysis versus recent predictions from simulations, we see our resulting value of $\sim$\,$25\,\mu$K to be broadly consistent with the value obtained by \citet{Baldi2018} for the MUSIC massive clusters with $M_{\rm vir}>5\times10^{14}h^{-1}M_{\odot}$, and at the same order of amplitude as the simulation results of \citet{Altamura2023} predictions using the MACSIS clusters. We caution that these comparisons should be taken as qualitative, due to the selection effects presented in our observational analysis, which differ from those in the simulations. Rigorous comparisons have to be made by applying the same selection procedure to those simulations, which is beyond the scope of this paper. 

Considering the significant differences between our observational selection and weighting scheme versus those recent simulation-based studies e.g., \citet{Baldi2018} and \citet{Altamura2023}, it would be difficult to make any direct quantitative comparison with them. Both studies focus on substantially more massive cluster-scale systems (with $M_{\rm vir}>5\times10^{14}h^{-1}M_{\odot}$), whereas our sample includes more lower-mass group-scale haloes. In addition, \citet{Baldi2018} further restrict their analysis to relaxed, high-spin clusters viewed close to edge-on, while both studies define the rotation axis directly from the intrinsic gas angular momentum in the simulations. In contrast, our analysis infers the dipolar axis observationally from the galaxy distribution, introducing additional projection and galaxy-gas misalignment effects. 
The study of \citet{Altamura2023} suggest that using galaxies as proxies for the gas rotation axis can suppress the recovered stacked amplitude of kSZ. But this is not directly relevant to our case, as our analyses with the simulation and observation would suffer the same effect. 
Perhaps coincidentally, they found an rkSZ amplitude of $12\pm 1.6\,\mu$K for their lowest-mass bin ($M_{500} < 9.7 \times 10^{14} M_{\odot}$ ), which is close to the simulated result inferred from ELUCID in our study. 
%As discussed in \citet{Altamura2023}, using galaxies as proxies for the gas rotation axis can suppress the recovered stacked amplitude by a factor of $\sim 2.5$. In particular, they also found an rkSZ amplitude of $12\pm 1.6\,\mu$K for their lowest-mass bin ($M_{500} < 9.7 \times 10^{14} M_{\odot}$ ) using the galaxy-aligned angular momentum direction, which is close to the simulated result inferred from ELUCID in our study. 
We caution any direct quantitative comparison between our measurement with those relatively idealised simulation results, despite their apparent coincidence in the same order of magnitude. %here remains broadly consistent in order of magnitude with the amplitudes reported in those simulation studies. Nevertheless, 
A rigorous quantitative comparison would require applying an observationally equivalent selection and axis-reconstruction procedure directly to the simulations, which is beyond the scope of this work.

During the preparation of this manuscript, another possible detection of the rkSZ signal was reported in \cite{Goldstein2025}. It is striking that the results seems to be consistent, despite the significant difference between our sample and the one used by \cite{Goldstein2025} -- 25 X-ray selected nearby clusters versus our 134 optically selected groups -- and our rather different analysis procedures. 

\section*{Acknowledgements}
YC acknowledges the support of the UK Royal Society through a University Research Fellowship. WC gratefully thanks Comunidad de Madrid for the Atracci\'{o}n de Talento fellowship no. 2020-T1/TIC19882 and Agencia Estatal de Investigación (AEI) for the Consolidación Investigadora Grant CNS2024-154838. He further acknowledges the Project PID2024-156100NB-C21 financed by MICIU/AEI /10.13039/501100011033/FEDER, EU and ERC: HORIZON-TMA-MSCA-SE for supporting the LACEGAL-III (Latin American Chinese European Galaxy Formation Network) project with grant number 101086388 and the science research grants from the China Manned Space Project. HYW is supported by the National Natural Science Foundation of China (NSFC, Nos. 12595312, 12192224), the CAS Project for Young Scientists in Basic Research, Grant No. YSBR-062 and the New Cornerstone Science Foundation through the XPLORER PRIZE. YC is grateful for the hospitality
of the Astrophysics and Theoretical Physics groups of the Department of Physics at the Norwegian University of Science and
Technology, and the School of Physics and Astronomy at Sun Yat-Sen University in China, during his visit, when part of this work was conducted. YC thanks Jens Chluba, Longlong Feng, Qi Guo, Adam Hincks, Xiaodong Li, Ian McCarthy, Kai Wang, Xin Wang, Dandan Xu, and Yi Zheng for their useful discussions.

For the purpose of open access, the author has applied a Creative Commons Attribution (CC BY) license to any author-accepted-manuscript version arising from this submission.

Many of the results in this paper have been derived using the \textsc{healpy} packages.

We thank the anonymous referee for providing detailed and constructive comments, which have helped to improve the presentation of the paper.

\section*{Data availability}
The group catalogue used in this analysis is publicly available at \url{https://gax.sjtu.edu.cn/data/Group.html}. The SDSS-DR7 galaxy sample is publicly available at \url{https://classic.sdss.org/dr7}. The ELUCID simulation is publicly available at \url{https://gax.sjtu.edu.cn/data/ELUCID.html}. The {\it Planck} temperature map and its mask are publicly available at \url{https://pla.esac.esa.int/pla}. Our analysis software is available upon reasonable request.

\bibliographystyle{mnras}
\bibliography{draft}

\end{document}